\documentclass{article}[10pt]

\usepackage[english]{babel}

\usepackage[font=footnotesize,labelfont=normalfont]{caption}
\usepackage{amsmath}
\usepackage{graphicx}
\usepackage{hyperref}
\usepackage{listings}
\usepackage{booktabs}
\usepackage{enumitem}
\usepackage{rotating}

\usepackage{flushend}

\title{Kernel Launcher: C++ Library for Optimal-Performance Portable CUDA Applications}
\author{Stijn Heldens, Ben van Werkhoven \\ 
\texttt{\{s.heldens,b.vanwerkhoven\}@esciencecenter.nl}\\
Netherlands eScience Center}
\date{\today}
\lstdefinestyle{mystyle}{
    language=C++,
    basicstyle=\ttfamily\footnotesize,
    breakatwhitespace=false,         
    breaklines=true,                 
    captionpos=b,
    escapechar=|,
    keepspaces=true,                 
    numbers=left,                    
    numbersep=5pt,                  
    showspaces=false,                
    showstringspaces=false,
    showtabs=false,                  
    tabsize=2
}

\begin{document}
\maketitle

% force page numbers
\thispagestyle{plain}
\pagestyle{plain}

\begin{abstract}

\emph{Graphic Processing Units} (GPUs) have become ubiquitous in scientific computing.
However, writing efficient GPU kernels can be challenging due to the need for careful code tuning.
To automatically explore the kernel optimization space, several auto-tuning tools -- like Kernel Tuner -- have been proposed. 
Unfortunately, these existing auto-tuning tools do not concern themselves with integration of tuning results back into applications, which puts a significant implementation and maintenance burden on application developers.
In this work, we present \emph{Kernel Launcher}: an easy-to-use C++ library that simplifies the creation of highly-tuned CUDA applications. With Kernel Launcher, programmers can \emph{capture} kernel launches, \emph{tune} the captured kernels for different setups, and \emph{integrate} the tuning results back into applications using runtime compilation. 
To showcase the applicability of Kernel Launcher, we consider a real-world computational fluid dynamics code and tune its kernels for different GPUs, input domains, and precisions.

\iffalse
have become ubiquitous in scientific computing. %, and many HPC systems rely on the massive computational power of these devices.
Writing efficient GPU code is considered challenging, since obtaining optimal performance requires careful code tuning.
Auto-tuning tools, like Kernel Tuner, are designed to explore the kernel optimization space automatically.
However, existing auto-tuning tools are limited in their capability to create optimal-performance portable applications, putting a significant implementation and maintenance burden on application developers.
In this work, we : An easy-to-use C++ library that simplifies the creation of highly-tuned CUDA applications.
Kernel Launcher enables programmers to
% unify the code to launch kernels with the code defining the tunable kernels
\emph{capture} kernel launches, 
\emph{tune} the captured kernels for different setups, 
and \emph{integrate} the tuning results back into applications using \emph{runtime compilation}.
%
We show that Kernel Launcher can be used to create performance-portable CUDA applications by considering a real-world computational fluid dynamics code and tuning two kernels for different GPUs, input domains, and precision.
\fi

% Stijn: We need more concrete results, like x% speedup in y% of the cases
% Ben: Dat is nou net lastig want Kernel Launcher geeft altijd 100% performance, je kan hooguit laten zien dat er beperkingen zijn als je minder goed tuned

\end{abstract}

\section{Introduction}
% Intro on GPUs
Computer systems used for \emph{High-Performance Computing} (HPC) are becoming more complex~\cite{heldens2020landscape}, making it increasingly more difficult to achieve good performance. 
This complexity is due in part to the widespread use of \emph{Graphics Processing Units} (GPUs). 
For instance, in November 2022, seven of the top ten systems in the TOP500~\cite{top500} achieved their performance solely thanks to GPUs.

% why GPU programming is challenging and requires 
Programming and optimizing applications for GPUs is generally considered to be more challenging than for CPUs~\cite{werkhoven2020lessons}.
One of the key challenges of GPU programming is the need to write GPU-specific functions, called \emph{kernels}, that are executed by millions of threads in parallel.
Deciding how to divide and assign work to threads is critical to achieving optimal performance. 
Additionally, the complex memory hierarchy of GPUs means that the decision on where and how to store data significantly impact performance.
Furthermore, optimizing code for GPUs involves trade-offs: certain code optimization may improve performance in some cases, but could be detrimental in others.
For example, loop unrolling, spatial blocking, prefetching and vectorization can all affect performance in different ways. 
We refer to the work by Hijma et al~\cite{hijma2022} for an extensive overview of GPU code optimizations discovered over the last 14 years.

% introduce Kernel Tuner
Optimizing GPU programs thus requires identifying performance-affecting parameters and tuning these parameters to achieve optimal performance.
Although tuning can be done manually, \emph{auto-tuning} tools can also be used to search for the optimal kernel configuration in an automated way.
%While this tuning could be done manually, \emph{auto-tuning} tools can also be used to search for the optimal kernel configuration in an automated way.
One such tool is \emph{Kernel Tuner}~\cite{kerneltuner}, an easy-to-use tool for tuning GPU code that support many search optimization methods.
With Kernel Tuner, GPU programmers create simple Python scripts that specify the kernel code, the tunable parameters, and the input data.
The tool then finds the best-performing kernel configuration by exploring the optimization search space.

% Introduce perf portability
However, these kernel configurations are often not portable since performance heavily depends on the specific input problem (e.g., dimensions or precision of input data) and the exact GPU architecture.
The kernel configuration that gives optimal performance for one particular workload or GPU may result in poor performance for a different setup.
For example, allowing threads to process more than one item could improve performance for larger datasets, but it limits concurrency for smaller datasets. Similarly, unrolling loops may be preferred for GPUs with large register files, as it maximizes data reuse in registers and increases instruction-level parallelism, but it results in excessive register spilling on GPUs with smaller register files.

In practice, GPU programmers may tune a CUDA kernel for a single GPU and input problem, and then falsely assume that the optimal configuration is sufficiently portable to all possible inputs and GPUs.  
Performance-portable GPU application would require tuning for every possible combination of input data set and GPU an integrating this knowledge into the application.
%achieving optimal performance for GPU applications thus requires tuning for every possible combination of input data set and GPU, complicating the creation of performance-portable GPU applications.
Compile-time kernel selection could be used, but this puts a maintenance burden on the application developers.
%Compile-time kernel selection based on the GPU may be applied, but it is difficult to implement when performance changes drastically for different input problems.
%In this paper, we show that these assumptions lead to suboptimal performance that is sometimes worse than not tuning at all.  % Stijn: Already in bullet list 
Ideally, it should be possible to tune kernels for many different setups and the application should then be able to select the optimal kernel automatically based on the situation at hand.
%However, support for this in current auto-tuning tools is limited and, therefore, difficult to implement.

% In this work
To address this issue, we present \emph{Kernel Launcher}: A C++ library that simplifies the integration of auto-tuning in CUDA applications by performing offline tuning of GPU kernels using Kernel Tuner.
We make the following contributions:
\begin{itemize} %[leftmargin=3pt]
  \item We introduce an API for tunable kernel definitions that unifies the host code to launch the kernel with the description of the kernel's tunable parameters.
  \item We introduce the concept of kernel {\em captures} that automatically export all information required to execute a kernel.
  \item We fully automate the use of Kernel Tuner for CUDA applications by \emph{replaying} the captured launches for different parameter configurations.  % , which can now tune kernels
  \item We demonstrate that only tuning for one dataset or one GPU (even of the same vendor and architecture) is insufficient to achieve optimal performance in all scenarios. Similarly, we show how tuning a kernel for only one particular scenario may result in worse performance compared to not tuning the code at all for other scenarios.
  \item Finally, we show that Kernel Launcher's runtime kernel selection and compilation can be used to create optimal-performance portable CUDA applications. 
\end{itemize}

%This paper is structured as follows.
%First, we provide background information on Kernel Tuner and auto-tuning in general.
%Next, we describe the implementation and features of our library Kernel Launcher.
%Afterwards, we demonstrate the benefits of using Kernel Launcher using X kernels from a real-world computational fluid dynamics code.
%Finally, we present conclusions and ideas for future work.
\section{Related Work}\label{sec:relatedwork}

We distinguish between two different research topics in the field of auto-tuning research. Some auto-tuning research focuses on (1) auto-tuning compiler-generated code optimizations~\cite{tiwari2009scalable,puschel2005spiral, SRTuner, compilePDEtune}, whereas this paper focuses on (2) software auto-tuning~\cite{li2009note,zhang2012auto}. Ashouri et al.~\cite{ashouri2018survey} wrote an excellent survey on machine-learning methods for compiler-based auto-tuning. 
In this paper, we focus on (2) auto-tuning software on the source code level, sometimes called \emph{automated design space exploration}~\cite{nardi2019hypermapper}.
Software auto-tuning is not limited to conservative assumptions that a compiler can make and, as such, allows to automatically optimize computational functions in isolation.
For example, it is possible to automatically select among many different optimization techniques~\cite{hijma2022} or entirely different implementations of the same algorithm.
Software auto-tuning is commonly employed by highly-optimized libraries and CPU applications (e.g., ATLAS~\cite{whaley1998automatically} and FFTW~\cite{frigo2005design}) as well as GPU applications~\cite{grewe2011automatically,li2009note,tomov2010dense,zhang2012auto,mametjanov2012autotuning,vanWerkhoven2014optimizing,sclocco2014auto}.
However, the technology used to tune these applications is application specific. 

Several generic auto-tuning frameworks have been introduced in recent years to
 create a reusable application-independent solution. 
In this way, advances in auto-tuning implementations can be reused for new projects and new results in auto-tuning research no longer have to be implemented in multiple application-specific tuners.
%
%generic auto-tuning frameworks, with a focus on optimization methods
%
\emph{OpenTuner}~\cite{ansel_opentuner_2014} was one of the first generic software auto-tuning frameworks, but with no support for tuning individual GPU kernels. 
\emph{GPTune}~\cite{liu2021gptune} is a tuning framework for MPI-based applications with support for parallel tuning.
\emph{HyperMapper}~\cite{nardi2019hypermapper} is a tuning framework for multi-objective optimization on FPGAs %using user-prior knowledge of FPGAs.
% mention ytopt as well

% GPU auto-tuning frameworks
\emph{CLTune}~\cite{nugteren2015cltune} was the first generic auto-tuning framework with support specifically for tuning OpenCL kernels, it has been used to implement an auto-tuned BLAS library implemented in OpenCL~\cite{nugteren2018clblast}. \emph{Auto-Tuning Framework} (ATF)~\cite{raschTACO} generates auto-tuning search spaces efficiently using a chain-of-trees search space structure.
\emph{Kernel Tuning Toolkit} (KTT)~\cite{filipovivc2017autotuning} is a C++ auto-tuning framework explicitly developed  to support online auto-tuning and pipeline tuning, which allows exploring of combinations of tunable parameters over multiple kernels. Online auto-tuning is a powerful technique that requires substantial modification of the host application, but allows applications to be tuned at runtime, during normal execution of the application. Online auto-tuning can be advantageous when performance strongly depends on input data. In this paper, we take an orthogonal approach, where we tune offline, once for each problem, and store the obtained performance results to disk.

In earlier work, we have introduced \emph{Kernel Tuner}~\cite{kerneltuner}, a compile-time auto-tuning framework that supports many different optimization algorithms~\cite{schoonhoven2022benchmarking}, including Bayesian Optimization~\cite{willemsen2021bayesian}. The compile-time auto-tuning approach taken by Kernel Tuner, allows auto-tuned kernels to be integrated into applications with minimal modification to the host application. Since the auto-tuning process happens at compile-time, there are no limitations on the programming language of the host application. 

% papers specifically focusing on run-time kernel selection
Falch and Elster~\cite{falch2017machine} have used machine learning for run-time kernel selection of OpenCL applications to use auto-tuning with to create performance portable applications. A neural network is trained using benchmark data, which is, in turn, used to select parts of the search space that are explored exhaustively. By building on top of Kernel Tuner, Kernel Launcher simply uses the results obtained by one of the optimization strategies implemented in Kernel Tuner.

% definitely need to cite this one https://arxiv.org/pdf/2008.13145.pdf
Lawson~\cite{lawson2020towards,lawson2021performance} uses unsupervised clustering to pre-select a subset of the compiled kernel configurations and uses a classification method to select from the best configuration at runtime. Kernel Launcher uses runtime compilation and, therefore, does not include pre-compiled kernel configurations with its applications, as there is no need to pre-select configurations.

Somewhat related to our framework is CERE~\cite{10.1145/2724717}, a framework for code isolation that can extract ``hotspots'' of an application as isolated code fragments for tuning. 
However, CERE explicitly targets LLVM code generation for CPUs, while Kernel Launcher is aimed at GPU kernels.

%BOAST~\cite{doi:10.1177/1094342017718068}

\section{Background}

This section introduces GPU auto-tuning with Kernel Tuner and explains how Kernel Launcher improves the software engineering experience by extending Kernel Tuner's capabilities.

Kernel Tuner is designed for tuning GPU kernels -- written in CUDA or OpenCL -- that are usually extracted from a larger application. % in any host programming language. 
While Kernel Tuner is written in Python, no code that uses Kernel Tuner needs to become part of the host application, nor does Kernel Tuner insert any dependencies into the kernel source code, which can still be compiled with their regular compilers after tuning.
This means that the host application can be written in any programming language.

To tune a kernel (see Listing~\ref{fig:vector-add-kernel}), application developers create a small Python script (see Listing~\ref{fig:kernel-tuner-example}) that specifies the kernel source code, name, problem size, arguments, and tunable parameters:
\begin{itemize}
\item The kernel \emph{source code} is specified by pointing to a file that should be able to compile in isolation.
\item The kernel \emph{name} in the source code. The name can include tunable parameter as C++ template arguments.
\item The \emph{problem size} denotes the dimensions of problem domain of the GPU kernel. For most kernels, the problem size is related to the size of the input/output domain of the GPU kernel. Kernel Tuner uses the problem size to calculate the total number of threads blocks. %, while the size of thread blocks may vary. 
\item The kernel \emph{arguments} is a list of multi-dimensional arrays and scalar values. These arguments are passed as function arguments to the GPU kernel. In most cases, the tuning script generates random arrays for the input data and declares empty arrays for the output data.
\item The \emph{tunable parameters} are specified as named parameters with lists of possible values. The parameters are passed as compile-time constants to the kernel code. 
\end{itemize}

Kernel Tuner supports many additionally arguments to specify, for example, search space restrictions, output verification, user-defined metrics, or tuning objectives. 
We refer to the Kernel Tuner documentation\footnote{\url{http://kerneltuner.github.io}} for an extensive description of the function {\tt tune\_kernel}.

Kernel Tuner is designed to automatically explore the search space by compiling configuration-specific kernels and then running benchmarks on the GPU. 
However, as the number of tunable parameters increases, the search space generally grows exponentially, which can make it impractical to explore exhaustively. 
In order to address this challenge, Kernel Tuner incorporates several optimization strategies to optimize the auto-tuning process.

Once the optimal kernel configuration has been determined by the tuner, this information needs to be integrated back into the target application. 
Kernel Tuner supports \emph{compile-time} kernel selection through an API that can generate C header files.
These header files include multiple targets, one for each GPU, that allows a build system (e.g., Make or CMake) to select the optimal kernel configuration for a particular target GPU during compilation. 
In this way, the application can achieve optimal performance a particular GPU, as long as the target GPU is known at compile-time and performance does not depend on the input size.

\begin{figure}[t!]\centering
\begin{lstlisting}[caption={Example of a CUDA kernel for vector addition.},label={fig:vector-add-kernel}]
template <int block_size>
__global__ void vector_add(float *c, float *a, float *b, int n) {
    int i = blockIdx.x * block_size + threadIdx.x;
    if (i<n) {
        c[i] = a[i] + b[i];
    }
}
\end{lstlisting}\vspace{-10pt}
%\end{figure}
%\begin{figure}[t!]
\centering
\begin{lstlisting}[language=Python,caption={Example a Python script using Kernel Tuner to tune the vector add kernel from Listing~\ref{fig:vector-add-kernel}},label={fig:kernel-tuner-example}]
import numpy as np
from kernel_tuner import tune_kernel

def tune():
    kernel_name = "vector_add<block_size>"
    kernel_file = "vector_add.cu"

    size = 10000000

    a = np.random.randn(size).astype(np.float32)
    b = np.random.randn(size).astype(np.float32)
    c = np.zeros(size).astype(np.float32)
    n = np.int32(size)

    args = [c, a, b, n]

    tune_params = dict()
    tune_params["block_size"] = [32, 64, 128, 256]

    return tune_kernel(kernel_name, kernel_file, size, args, tune_params)
\end{lstlisting}\vspace{-20pt}
\end{figure}

There are three areas where Kernel Launcher aims to improve both the software engineering process and the performance portability of the resulting tuned application. 
First, currently, users need to create a separate Python script for each kernel that requires tuning. 
While this approach may work for applications with only a few bottleneck kernels, it does not scale well and becomes unmanageable for large applications that have dozens of kernels.

Second, Kernel Tuner requires the user to generate valid input data to be used by the kernel.
%'s interface is written in Python and expects the kernel arguments to be specified using Python data types, such as Numpy, Cupy, or Torch ndarrays
This works well for kernels that use simple data structures or accept randomly generated data. 
However, for kernels that operate on complex data structures, like lookup tables or graphs, generating realistic input data that matches the data used in real runs of the application can be challenging. 
This places a significant burden on application developers.

Third, while the compile-time kernel selection functionality offered by Kernel Tuner has the advantage that little to no modification are required to the host application, it has some clear limitations.
To achieve optimal performance on every GPU and problem size, the application needs to be recompiled every time. 
Additionally, using a compile-time kernel selection approach in applications where the same kernel may be executed on different problems within a single run is complex and requires considerable engineering effort.

Kernel Launcher aims to enhance the software engineering experience of using Kernel Tuner for developing tunable applications. 
It goes beyond the existing capabilities of Kernel Tuner by offering runtime kernel selection and compilation. 
With these features, Kernel Launcher enables the creation of optimal, performance-portable applications that can reuse the same tunable kernel for different problems within a single execution of the same application.
\section{Kernel Launcher}

\begin{figure}
    \centering
    \includegraphics[width=\columnwidth]{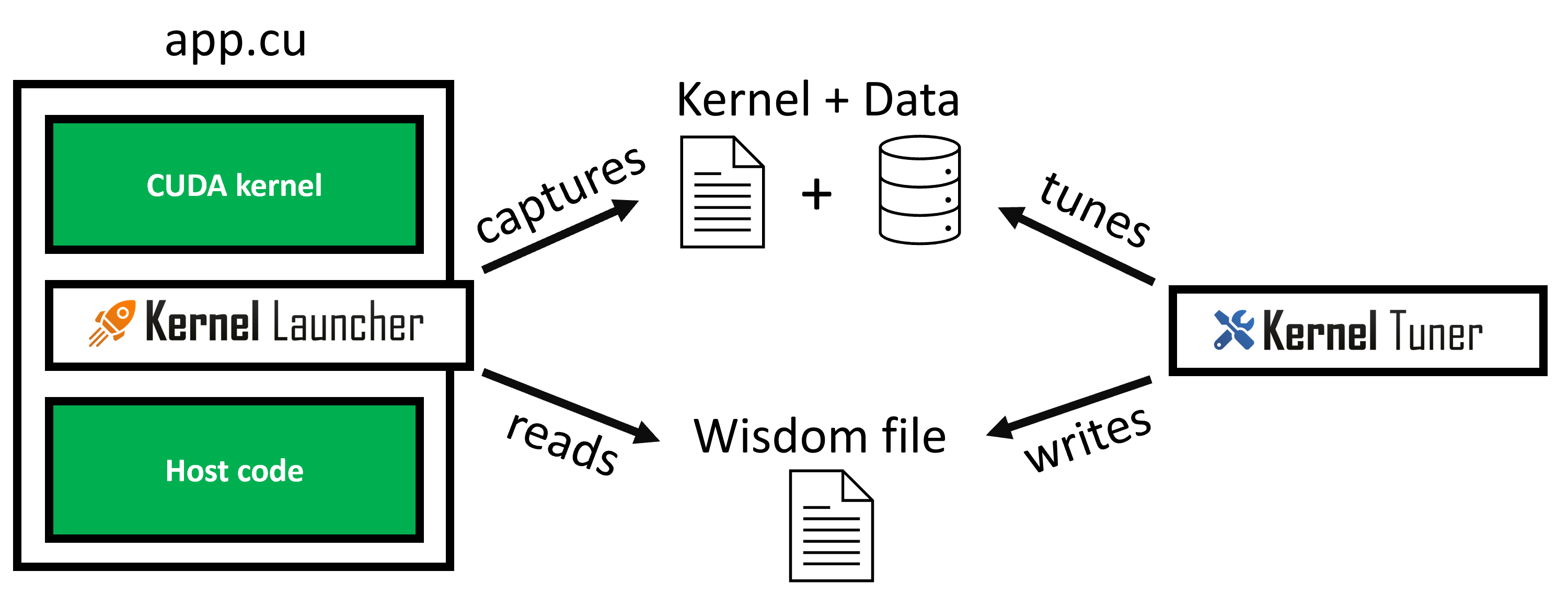}
    \caption{Overview of the designed interaction between the host application, Kernel Launcher, and Kernel Tuner.}\label{fig:kernel-launcher}
\end{figure}

In this section, we describe the implementation of \emph{Kernel Launcher}, which is implemented as a C++ library.
Figure~\ref{fig:kernel-launcher} shows how the library is integrated into applications and interacts with Kernel Tuner. 
The remainder of this section explains how Kernel Launcher is utilized within applications, and outlines each of the four steps shown in Figure~\ref{fig:kernel-launcher}. 
First, the user writes the \emph{kernel definition} using Kernel Launcher in C++ to allows kernels to be \emph{captured}. 
Then, Kernel Tuner tunes the kernels, producing wisdom files. These files allow Kernel Launcher to perform \emph{dynamic kernel selection} using \emph{runtime kernel compilation}.

\subsection{Kernel Definition}

To make a CUDA kernel tunable, the programmer defines its specifications using the Kernel Launcher API. 
These specifications consist of three elements: The configuration space, the compilation specifications, and the launch parameters.

For the configuration space, the definition includes the tunable parameters, the allowable values for those parameters, and any constraints on the search space (i.e., boolean expressions).
For the compilation specifications, the programmer must provide details such as the source code, kernel name, compiler flags, template arguments, and preprocessor definitions. 
To launch the kernel, the programmer must specify how the thread block size, the number of thread blocks, and the amount of shared memory are derived from the kernel arguments.

Once a kernel has been defined, it can be launched using the Kernel Launcher API by providing the kernel arguments in way that is similar to launching a regular kernel in plain CUDA (See Listing~\ref{fig:kernel-launcher-example},Line~\ref{line:launch}).
If the kernel has not yet been tuned, the default values for the tunable parameters are used.

Kernel Launcher thus consolidates the description of the tunable aspects of kernels and the code for launching the kernel, eliminating duplication between the Kernel Tuner script and hos code of application and moving this definition into the source code of the host application. 
Previously, with Kernel Tuner, the definition of the tunable kernel and its parameters resided in separate Python files, and the relationship between the problem size and tunable parameters was described in both the Kernel Tuner script and the host application.
This distribution of the kernel definition across multiple files led increased maintenance costs since all files need to be kept up to date when changes are made to the kernel source code. 
However, with Kernel Launcher, the kernel definition and its launch code are integrated in the source code of the host application, resulting in significant maintenance cost savings for C++ applications with many tunable kernels.

\subsection{Kernel Capturing}
An important concept in Kernel Launcher is the ability to \emph{capture} a kernel launch, which enables storage of all information necessary to execute the kernel, including the kernel definition and possible the input data.
By capturing a kernel, auto-tuning tools can `replay' the exact same kernel launch for different configurations of the tunable parameters. 
%This capture mechanism ensures that the kernel execution uses the same data as was used by the application at runtime.

There are two main advantages of this capturing concept.
First, it removes the need for programmers to have to generate input data when tuning kernels and, instead, making it possible to tune the kernel directly on real-world data extracted from the application. 
This approach can be particularly useful for complex data sets that are difficult to generate off-line.
%Instead, the kernel can be tuned directly on real-world data extracted from the application.
Second, it enables offline tuning instead of having to tune at runtime.
Offline tuning means that more accurate measurements can be obtained since the kernel is executed in isolation.
In contract, runtime tuning could lead to inconsistent measurement if, for example, the GPU also executing other kernels concurrently or the CPU is overloaded.

To capture kernel launches in our implementation, the programmer needs to set the environment variable \verb=KERNEL_LAUNCHER_CAPTURE= to a list of kernel names separated by commas.
It is possible to capture multiple kernels in a single run of the application.

\subsection{Kernel Tuning}

After capturing a kernel launch, the associated kernel can be tuned.
Kernel Launcher includes a command-line script that, in turn, uses Kernel Tuner~\cite{kerneltuner} as the main tool for auto-tuning, although in principle, other GPU auto-tuners, such as KTT~\cite{filipovivc2022using}, could be used as well.

Several command-line options can be provided to the script, such the search strategy and the termination condition. 
The default setting is to tune each kernel for at most 15 minutes using the Bayesian Optimization search strategy~\cite{willemsen2021bayesian}.

\subsection{Wisdom Files}
After the auto-tuner terminates, the best-performing configuration is written to a human-readable file which we referred to as \emph{wisdom files}.
The term ``wisdom'' was originally introduced by FFTW~\cite{frigo1998fftw}, but Kernel Launcher borrows only the terminology and uses a different file format.

For each kernel in the application, a corresponding wisdom file is available that stores the results of all tuning sessions for that specific kernel.
Re-tuning the same kernel multiple times for various problem sizes or GPUs add new results to the existing wisdom file.

A wisdom file consists of a sequence of records.
Each record represents the best-performing configuration found during auto-tuning session for one particular GPU and \emph{problem size}.
The problem size is a multi-dimensional vector that indicates the size of the workload.
The interpretation of the problem size varies depending on the specific problem and is defined as part of the kernel definition.
For example, for matrix multiplication of matrices of sizes $n{\times}m$ and $m{\times}k$, the problem size would be $(n, k, m)$.
In addition to the tuning results, each record includes provenance information associated with the tuning sessions, such as the date, software versions, GPU properties, and the host name.

\subsection{Kernel Selection and Compilation}
During application execution,
Kernel Launcher selects the optimal configuration for each combination of kernel, GPU, and problem size.
The first time a kernel is launched, Kernel Launcher processes the wisdom file for that kernel and selects one record based on the GPU and problem size, using the following heuristic:

\begin{itemize}
\item
If a record exists that matches the GPU and the problem size, that record is chosen.

\item
If no such record exists, the record that matches the GPU and problem size that is closest in Euclidian distance.

\item
If no record exists that matches the current GPU, the record that matches the GPU \emph{architecture} and the problem size is closest is chosen.

\item
If no record exists that matches the current GPU architecture, the record that has the closest problem size is chosen.

\item
If the wisdom file is empty or missing, the default configuration is selected.
\end{itemize}

% TODO: Stijn: NVRTC needs reference?
After selecting a configuration, Kernel Launcher compiles the kernel code at runtime by NVRTC, the NVIDIA runtime compilation library, and loads the compiled code onto the GPU. 
%As part of the run-time compilation, the code for the selected kernel configuration is generated with the values of all tunable parameters inserted in the code as compile-time constants. 
Note that kernel selection and compilation only happens on the first launch of a kernel for a given problem size and on subsequent calls for the same problem size will reuse the compiled instance of the kernel configuration.

\begin{figure}[t!]
\begin{lstlisting}[caption={Example of a kernel definition in Kernel Launcher},label={fig:kernel-launcher-example}]
#include <kernel_launcher.h>

void run(float *c, float *a, float *b, int n) {
    // create builder.
    auto builder = kernel_launcher::KernelBuilder(
        "vector_add", "vector_add.cu");  
    auto block_size = builder.tune("block_size", 
        {32, 64, 128, 256, 1024});
    
    builder
        .problem_size(kl::arg3)
        .template_args(block_size)
        .block_size(block_size);

    // Create kernel
    auto kernel = kernel_launcher::WisdomKernel(
        builder);

    // Launch kernel
    kernel.launch(c, a, b, n);|\label{line:launch}|
}
\end{lstlisting}\vspace{-10pt}
\end{figure}

\subsection{Code Example}

Kernel Launcher is implemented as an easy-to-use C++ library. Figure~\ref{fig:kernel-launcher-example} shows an example of the code needed to integrate a tunable vector addition kernel written in CUDA into a C++ host application using Kernel Launcher. 
First, a \texttt{KernelBuilder} is instantiated that specifies the kernel name and the source file that contains the tunable CUDA kernel code. 
Second, the tunable parameters and valid values are specified. 
Then, the problem size, template arguments and thread block dimensions are specified. Note that these may be specified using the kernel arguments or tunable parameters.
Next, a \texttt{WisdomKernel} object is instantiated, which searchers the wisdom file and prepares the  kernel for runtime compilation. 
Finally, the kernel is launched on line~\ref{line:launch}.
Note that the thread block and grid dimensions are calculated by Kernel Launcher and should not be passed by the user.

% Ben: which commands do I need to type in to tune the kernels to create wisdom files? Can we do this from a Make or CMake file?

\section{Experimental Evaluation}
In this section, we present an experimental evaluation of Kernel Launcher.
Experiments were performed on the VU site of the DAS6 distributed supercomputer~\cite{bal2016medium}.
Table~\ref{tab:gpu-properties} lists the properties of the GPUs used for the evaluation.
Software versions were CUDA toolkit 11.7, CUDA driver 515.65.01, Kernel Tuner 0.4.3, and CentOS 8.5.

\begin{table}[]
    \scriptsize
        \caption{GPUs used in our experiments. Bandwidth (BW) in GB/s. Peak single precision (SP) and double precision (DP) performance in GFLOP/s.}\label{tab:gpu-properties}
    \centering
    \begin{tabular}{l|l|r|r|r|r}
    \toprule
    GPU & Architecture & Cores & BW & Peak SP & Peak DP \\
    \midrule
    RTX A4000 & Ampere (GA104) & 6{,}144 & 448 & $19{,}170$ & 599 \\
    Tesla A100 & Ampere (GA100) & 6{,}912 & 1{,}555 & $19{,}500$ & 9700 \\
    \bottomrule
    \end{tabular}
  \vspace{-.3cm}
\end{table}

\subsection{Application}

To evaluate the usability and performance of Kernel Launcher, we have integrated our library into % to tune several kernels.
MicroHH~\cite{gmd-10-3145-2017,chiel_van_heerwaarden_2017_822842}, a computational fluid dynamics software for direct numerical simulation and large-eddy simulation of turbulent flows in the atmospheric boundary layer.
This C++ application runs on both multi-core CPUs and CUDA-enabled GPUs.
The code supports many different integration schemes and nemerous configuration options, including the size of the three-dimensional domain grid and floating-point precision (single or double precision).
MicroHH is an excellent candidate for runtime kernel selection since achieving high performance requires distinct tuning parameters for each combination of GPU architecture, grid size, and floating-point precision.
% Stijn: Interesting candidate?
% Ben: tja, eigenlijk zou je willen dat je kunt stellen dat MicroHH regelmatig op verschillende GPUs of probleem groottes gedraaid wordt (het liefst zelfs binnen dezelfde run), want dan heb je pas echt een argument voor run-time kernel selectie. Je kunt als je het bovenstaande argument niet zo sterk vind het ook gewoon weglaten. We hebben voor de evaluatie microhh gebruikt, punt. Als je het niet noemt is er vaak ook niemand die er een punt van maakt.

\subsection{Tunable Parameters}
For this paper, we selected two kernels from MicroHH for analysis.
The first kernel is called \verb=advec_u= and corresponds to a large stencil operation.
In particular, the kernel performs advection along the X-axis and is part of the second order advection scheme with fifth order interpolation.
The second kernel is called \verb=diff_uvw= and is an element-wise operation.
This kernel is part of the second order Smagorinsky diffusion for large-eddy simulation.

Both kernels work on a three-dimensional grid where each grid point requires an operation to be performed.
the kernels launched a single CUDA thread for each grid point, with each thread processing the point associated with its 3D global thread index.
However, we have rewritten the kernels and introduced several code optimizations that resulted in the following tunable parameters (see Table~\ref{tab:evaluation_params}):

\begin{table}
\caption{List of tunable parameters and their default value.}
\label{tab:evaluation_params}
\centering
\begin{tabular}{l||p{5cm}|l}
\toprule
Name & Values & Default value \\
\midrule
Block size X & 16, 32, 64, 128, 256 & 256 \\
Block size Y & 1, 2, 4, 8, 16 & 1 \\
Block size Z & 1, 2, 4, 8, 16 & 1 \\
Tile factor X & 1, 2, 4 & 1 \\
Tile factor Y & 1, 2, 4 & 1 \\
Tile factor Z & 1, 2, 4 & 1 \\
Unroll X & \texttt{true}, \texttt{false} & \texttt{false} \\
Unroll Y & \texttt{true}, \texttt{false} & \texttt{false} \\
Unroll Z & \texttt{true}, \texttt{false} & \texttt{false} \\
Tile contiguous X & \texttt{true}, \texttt{false} & \texttt{false} \\
Tile contiguous Y & \texttt{true}, \texttt{false} & \texttt{false} \\
Tile contiguous Z & \texttt{true}, \texttt{false} & \texttt{false} \\
Unravel permutation & \texttt{XYZ}, \texttt{XZY}, \texttt{YXZ}, \raggedright \texttt{YZX}, \texttt{ZXY}, \texttt{ZYX} & \texttt{XYZ} \\
Min. blocks per SM & 1, 2, 3, 4, 5, 6 & 1 \\
\bottomrule
\end{tabular}
\vspace{-.2cm}
\end{table}

\begin{itemize}[leftmargin=*]
\item 
\textbf{Block Size XYZ}.
The number of threads in each CUDA thread block along the X, Y, and Z axis.
The default thread block size is $(256, 1, 1)$

\item
\textbf{Tiling Factor XYZ}.
We modified the kernel so  that each thread processes multiple grid points along the X, Y, and Z axis.
For example, tiling factors $(3, 1, 2)$ indicate that each thread processes 6 points: 3 along the X-axis and 2 along the Z-axis.
Processing multiple items per thread can increase cache utilization and reduce thread scheduling overhead, but assigning too many items to each thread results in insufficient concurrency.
By default, no tiling is used. %, i.e. $(1, 1, 1)$.

\item
\textbf{Loop Unrolling XYZ}.
The tiling involves three nested loops, one for each axis of the 3D grid.
Each of the three loops can either be \emph{fully} unrolled or \emph{not} unrolled.
Unrolling a loop increases instruction-level parallelism and could result in additional compiler optimizations, at the cost of increased register usage and instruction count. 
There is a boolean parameter for each axis.

\item
\textbf{Tiling stride XYZ}.
Since each thread processes multiple items, there are multiple ways to assign the grid points to threads.
We implemented two options for each axis.
The first option is that each threads processes the points that are consecutive along one axis (e.g., item $x, x+1, x+2, \ldots$).
The second option is that the points are block-strided (e.g., item $x, x+b_X, x+2b_X, \dots$ where $b_X$ is the block size along the X axis).
The first option can lead to better data reuse for stencil-like kernels (especially if the loop is unrolled), while the second option may give better memory access patterns.
%There is a boolean parameter for each axis.

\item
\textbf{Unravel permutation}.
The thread blocks are launched as a one-dimensional list of blocks.
Each thread then \emph{unravels} its 1D thread block identifier into a 3D grid index.
Since there are six possible permutations of $(X,Y,Z)$, there are six possible ways to perform this unraveling.
We added this parameter since it affects the scheduling order of thread blocks and thus impacts cache utilization on the GPU.
For example, for the unravel permutation $(Z,X,Y)$, the order in which thread blocks are scheduled is first along the Z axis, then X axis, and finally the Y axis.

\item
\textbf{Thread blocks per SM}.
The number of threads that can reside on a \emph{streaming multiprocessor} (SM) impacts performance.
However, there is a delicate trade-off between available concurrency and register usage:
More threads per SM means more concurrency, at the cost of fewer available registers per thread.
We use CUDA's \verb=__launch_bounds__= compiler hint to tune the number of thread blocks per SM, and thus the register usage.

\end{itemize}

\subsection{Capture Results}

\begin{table}[]
\caption{Time and size required to capture kernel.}
\label{tab:evaluation_captures}
\centering
\begin{tabular}{l|l|l||l|l}
\toprule
Kernel & Grid size & Precision & Capture time & Capture size \\
\midrule
%time=2331538 size=70842672
\texttt{advec\_u} & $256^3$ & float & 2.3\,sec & 70.8\,MB \\
%time=4621459 size=141685344
\texttt{advec\_u} & $256^3$ & double & 4.6\,sec & 141.7\,MB \\
%time=18248830 size=551678256
\texttt{advec\_u} & $512^3$ & float & 18.2\,sec & 551.6\,MB \\
%time=43263257 size=1103356512
\texttt{advec\_u} & $512^3$ & double & 43.2\,sec & 1103\,MB \\
%time=5576891 size=212798464
\texttt{diff\_uvw} & $256^3$ & float & 5.6\,sec & 212.8\,MB \\
%time=11928014 size=425596928
\texttt{diff\_uvw} & $256^3$ & double & 11.9\,sec & 425.6\,MB \\
%time=43341804 size=1656099840
\texttt{diff\_uvw} & $512^3$ & float & 43.3\,sec & 1656\,MB \\
%time=82349427 size=3312199680
\texttt{diff\_uvw} & $512^3$ & double & 82.3\,sec & 3312\,MB \\
\bottomrule
\end{tabular}
\end{table}

\begin{sidewaysfigure*}
\includegraphics[width=\textwidth]{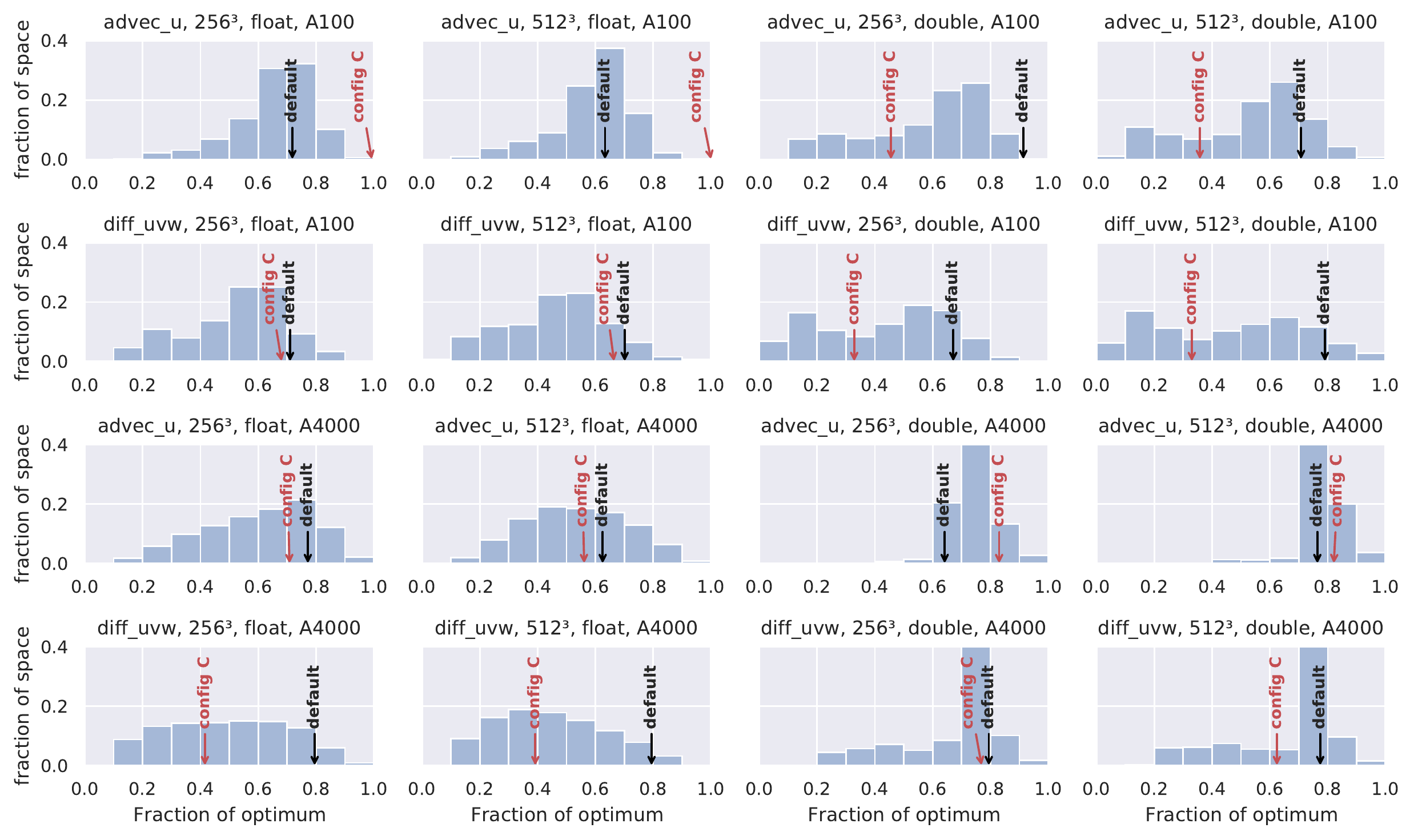}
\caption{Tuning results for each combination of kernel, grid size, floating-point precision, and GPU. Each graph is a histogram showing the distribution of performance for all kernel configurations in the entire search space. Performance is expressed as the runtime ratio compared to the optimal configuration for that particular scenario. The arrow at \emph{default} indicates the performance of the default kernel configuration. The arrow at \emph{configuration $\mathcal{C}$} indicates the performance of  the optimal configuration for \texttt{advec\_u}-$256^3$-float-A100 (The top left scenario).}
\label{fig:evaluation_histograms}
\end{sidewaysfigure*}

% Ben; Ik moest dit figure helemaal hier zetten om ervoor te zorgen dat het opzelfde pagina kwam als de tekst die dit figuur bespreekt
\begin{figure*}[t!]
\includegraphics[width=\textwidth]{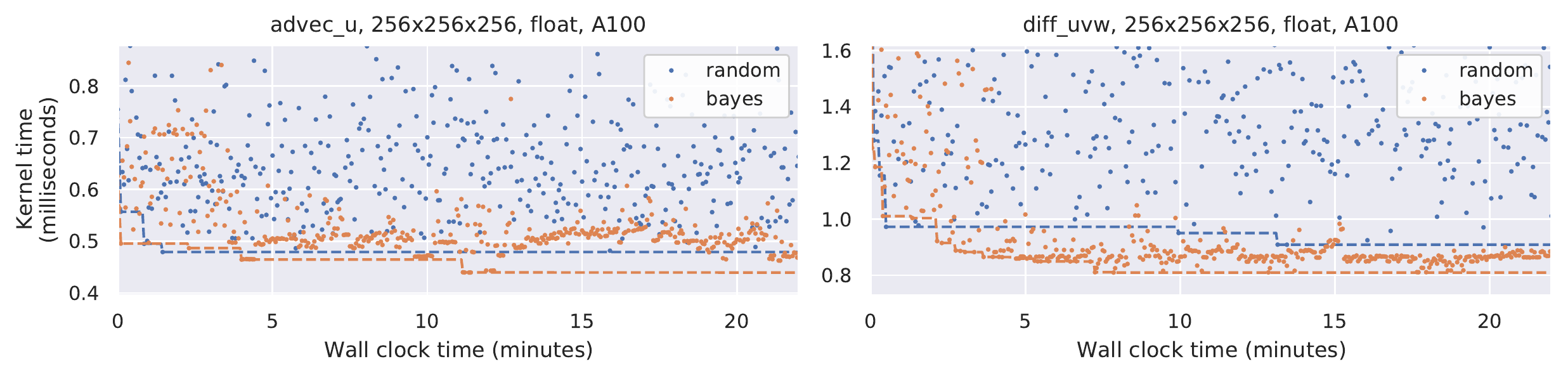}
\caption{Tuning sessions for both random and Bayesian optimization (\texttt{bayes}) search strategy. Each dot represents one configuration evaluated by the tuner. The vertical axis is the measured kernel time (note: does not start at zero). The horizontal axis is the wall clock time in minutes of the tuning session. The dashed lines indicates the performance of the the best configuration found during the tuning session.}
\label{fig:evaluation_tuning}
\end{figure*}

We executed MicroHH for two grid sizes ($256{\times}256{\times}256$ and $512{\times}512{\times}512$) and two floating-point precision (\texttt{float} and \texttt{double}).
Table~\ref{tab:evaluation_captures} shows the time required to capture each kernel and the size of the capture on disk.
The captures are written to persistent storage on a shared file system (NFS); thus,  IO times depend heavily on the load of the file system.
The table shows that the capture time required scales with the size of the capture which, in turn, scales with the grid size.

Considering all tunable parameters and their possible values, the entire search space consists of more than 7.7 million kernel configurations, which means exhaustively tuning each kernel in the application is infeasible. As such, we tune the eight captured kernels for two search strategies from Kernel Tuner: 
a \emph{random} search, since it gives an unbiased performance distribution, 
and \emph{Bayesian optimization}, since previous work~\cite{willemsen2021bayesian} shows it outperforms other search strategies.
Figure~\ref{fig:evaluation_tuning} shows two examples of the tuning sessions for a grid size of $256^3$, single precision, and an NVIDIA A100 GPU.
The other tuning sessions show similar result and are, therefore, not included.

These results show that both search strategies find increasingly better performing kernel configurations.
Since an exhaustive search is infeasible, we consider the best-performing configuration found after one hour of tuning to be the \emph{optimum}. 
Bayesian optimization strongly prefers better-performing kernel configurations, whereas the configurations selected by random search show a much larger spread. 
We found that Bayesian optimization takes, on average, 3.4 minutes (max: 6.5 minutes) to find a configuration 10\% away from the optimum and 7.5 minutes (max: 19 minutes) for a 5\% difference. 
A complete comparison of the different optimization strategies in Kernel Tuner is outside the scope of this work and we refer interested readers to the work by Schoonhoven et al.~\cite{schoonhoven2022benchmarking}.

% Stijn: Maybe we should say something about tuning times? Like, what is the average time to find the optimum?

\subsection{Tuning Results}

Next, we evaluate the tuning results of all eight captured kernels for the two GPUs: the A4000 and the A100.
In the remainder of this paper, we shall refer
to each combination of kernel, grid size, precision, and GPU as a \emph{scenario}. 
We will denote each scenario as \emph{kernel}-\emph{grid}-\emph{precision}-\emph{GPU} (e.g., \texttt{advec\_u}-$256^3$-{float}-{A100}).

\begin{sidewaysfigure*}
\includegraphics[width=\textwidth]{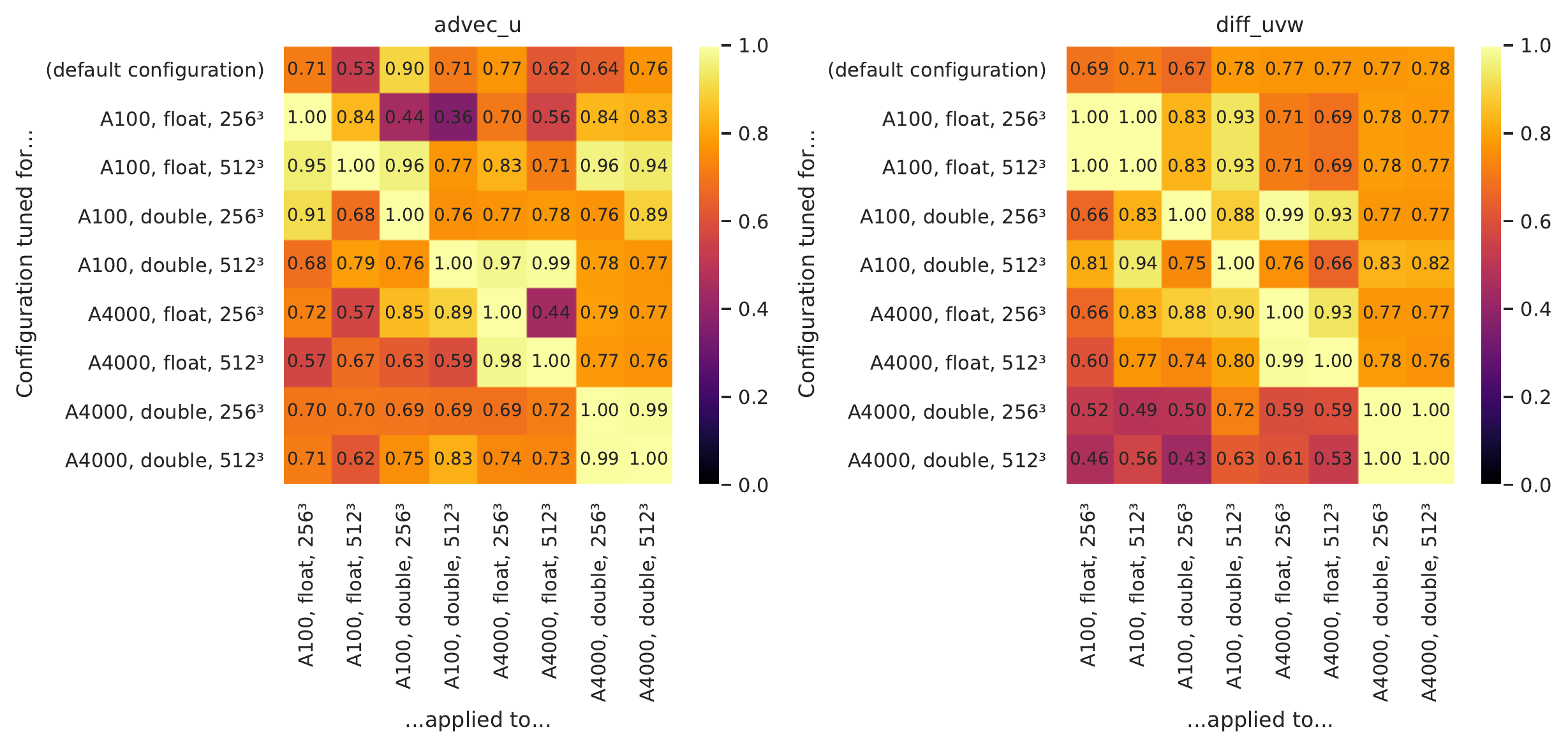}
\caption{Matrix showing how well the optimal configuration for one scenario performs in another scenario. The numbers are the fraction of optimum.}
\label{fig:evaluation_matrix}
\end{sidewaysfigure*}

Figure~\ref{fig:evaluation_histograms} shows 16 histograms of the kernel times measured by the random search strategy for each scenario.
The performance of each configuration is expressed as the fraction relative to the best performing configuration (i.e., the \emph{optimum}) for that particular scenario.
For example, a fraction of $0.8$ indicates that a configuration only reached 80\% of the performance compared to the optimum and is, thus, $1/0.8 = 1.25$ times slower.
The histogram shows how difficult it would be to tune these kernels by hand.
%These histograms give a good indication of how difficult it is to tune these kernels.
For example, for the scenario \texttt{advec\_u}-$256^3$-float-A100, only 0.8\% of the configurations are within 10\% of the optimum.
As such, it would be challenging and time-consuming to find this optimum manually, which is why auto-tuning tools are a necessity.

Each graph includes a \emph{black} arrow that indicates the performance of the default configuration.
The Figure shows that, for each graph, the default configuration is not near the optimum, meaning that auto-tuning can significantly increase performance. % for these kernels.
The default configuration across all 16 scenarios reaches only an average performance of 75\%, meaning tuning can increase performance by an average of 25\%.

We found that the optimal configuration is different for each of the 16 scenarios, and any particular configuration that gives high performance for one scenario usually performs poorly for others.
To illustrate this, we consider the best kernel configuration $\mathcal{C}$ for the scenario \texttt{advec\_u}-$256^3$-float-A100.
Figure~\ref{fig:evaluation_histograms} shows a \emph{red} arrow in each histogram that indicates the performance of $\mathcal{C}$ for each of the 16 scenarios.
The Figure shows that while $\mathcal{C}$ performs well in certain scenarios, it performs exceptionally poorly in other scenarios.
Configuration $\mathcal{C}$ performs worse than the default configuration in 11 of the 16 scenarios.
This result indicates that a kernel tuned for one particular scenario may not perform well in other scenarios, demonstrating the need to individually tune the kernels for all scenarios.

\subsection{Performance Portability}

To quantify the portability of the optimal configuration found when tuning from one scenario to another, Figure~\ref{fig:evaluation_matrix} shows the normalized relative performance. % of the ported configuration.

For example, when the optimal configuration found for 
\texttt{advec\_u}-$256^3$-float-A4000 is applied to the 
\texttt{advec\_u}-$512^3$-float-A4000 kernel, it achieves only 44\% of the best-known configuration for that problem. It is important to realize that between these two kernels, only the grid size has changed and both use single precision on the A4000. 

Conversely, for double precision, the optimal performance is very close for both $256^3$ and $512^3$. 
This is because of the design of the A4000, which has a limited number of double-precision \emph{floating-point units} (FPUs): only 1/32nd compared to the number of single-precision FPUs. Therefore, the kernels using double precision are compute-bound on the A4000, meaning many kernel configurations in the search space show similar performance once their memory-throughput is efficient enough to run into the bottleneck introduced by the limited number of double-precision FPUs. This observation is confirmed by the narrow search space distributions for kernels using double precision on the A4000. 

The A100 has many more double-precision FPUs, as its double-precision peak performance is half the single-precision peak performance. Consequently, we see more differences and, therefore, even less portability of the optimal configuration from one scenario to the next. For example, where the optima of \texttt{advec\_u}-$256^3$-double-A4000 and \texttt{advec\_u}-$512^3$-double-A4000 show near-identical performance, \texttt{advec\_u}-$256^3$-double-A100 and \texttt{advec\_u}-$512^3$-double-A100 only achieve 76\% of the optimal performance when applied to each other's scenario. 
The results for the \texttt{diff\_uvw} kernel are similar, except that varying only the problem size only leads to suboptimal performance for double precision on the A100.

\begin{table}[t]
    \caption{Performance portability metric (PPM) for the \texttt{advec\_u} kernel using the default or optimal configurations applied to each scenario. Kernel Launcher always selects the optimal configuration.}
    \label{tab:advec-u-ppm}
    \centering
    \begin{tabular}{l|c|c|c}
\toprule
Configuration tuned for	& Best	& Worst	& PPM \\
\midrule
(default configuration)	& 0.90	& 0.53	& 0.69 \\
A100, float, $256^3$	& 1.00	& 0.36	& 0.62 \\
A100, float, $512^3$	& 1.00	& 0.71	& 0.88 \\
A100, double, $256^3$	& 1.00	& 0.68	& 0.81 \\
A100, double, $512^3$	& 1.00	& 0.68	& 0.83 \\
A4000, float, $256^3$	& 1.00	& 0.44	& 0.71 \\
A4000, float, $512^3$	& 1.00	& 0.57	& 0.72 \\
A4000, double, $256^3$	& 1.00	& 0.69	& 0.75 \\
A4000, double, $512^3$	& 1.00	& 0.62	& 0.78 \\
Kernel Launcher	& 1.00	& 1.00	& 1.00 \\
\bottomrule
    \end{tabular}
\vspace{20pt}    
%\vspace{-8pt}
%\end{table}
%\begin{table}[t]
    \caption{Performance portability metric (PPM) for the \texttt{diff\_uvw} kernel using the default or optimal configurations applied to each scenario. Kernel Launcher always selects the optimal configuration.}
    \label{tab:diff-uvw-ppm}
    \centering
    \begin{tabular}{l|c|c|c}
\toprule
Configuration tuned for& Best	& Worst	& PPM \\
\midrule
(default configuration)	& 0.78	& 0.67	& 0.74 \\
A100, float, $256^3$	& 1.00	& 0.69	& 0.82 \\
A100, float, $512^3$	& 1.00	& 0.69	& 0.82 \\
A100, double, $256^3$	& 1.00	& 0.66	& 0.84 \\
A100, double, $512^3$	& 1.00	& 0.66	& 0.81 \\
A4000, float, $256^3$	& 1.00	& 0.66	& 0.83 \\
A4000, float, $512^3$	& 1.00	& 0.60	& 0.79 \\
A4000, double, $256^3$	& 1.00	& 0.49	& 0.63 \\
A4000, double, $512^3$	& 1.00	& 0.43	& 0.60 \\
Kernel Launcher	& 1.00	& 1.00	& 1.00 \\
\bottomrule
    \end{tabular}\vspace{-8pt}
\end{table}

Overall, we see that 
the configuration that is optimal for a given precision and problem size on the A100 only yields 70\%-77\% for \texttt{advec\_u} and 69\%-82\% for \texttt{diff\_uvw} of the performance that could have been achieved on A4000, if the optimal configuration for that scenario had been used instead. 
Similarly, using the optimal configurations for A4000 only achieves 67\%-83\% for \texttt{advec\_u} and 50\%-77\% for \texttt{diff\_uvw} of the performance that could have been achieved on A100.
This is surprising since both GPUs are from the same vendor and even belong to the same architecture (Ampere).
These results demonstrate the need for software development tools to use auto-tuning results obtained in different scenarios.

We computed the \emph{performance portability metric}~\cite{pennycook2016metric} (PPM) to quantify the portability of the default and optimal configurations to all scenarios, as shown in Tables~\ref{tab:advec-u-ppm} and \ref{tab:diff-uvw-ppm}. 
As can be seen in both tables, the performance portability score of the default configuration is in the same range as the scores for the configurations that have been tuned for one scenario. 
This means that tuning for only one scenario and then using this configuration in all scenarios may lead to worse overall performance compared to not tuning all. 
However, using Kernel Launcher's runtime kernel selection, the application always selects the optimal configuration to achieve the 100\% of the potential performance in each scenario.
To create performance portable applications that can achieve optimal performance in every scenario, it is necessary to select between different configurations based on the problem at hand.
%These results also highlight the limitations of Kernel Tuner.  % Stijn: Reviewer 2 does not like this sentence
Without Kernel Launcher's runtime kernel selection and runtime kernel compilation capabilities, a similar result could only have been achieved by the user recompiling the application every time it is executed on a different problem.

\subsection{Kernel Launcher Overhead}
\label{sec:evaluation_overhead}
\begin{figure}
\centering
\includegraphics[width=.5\textwidth]{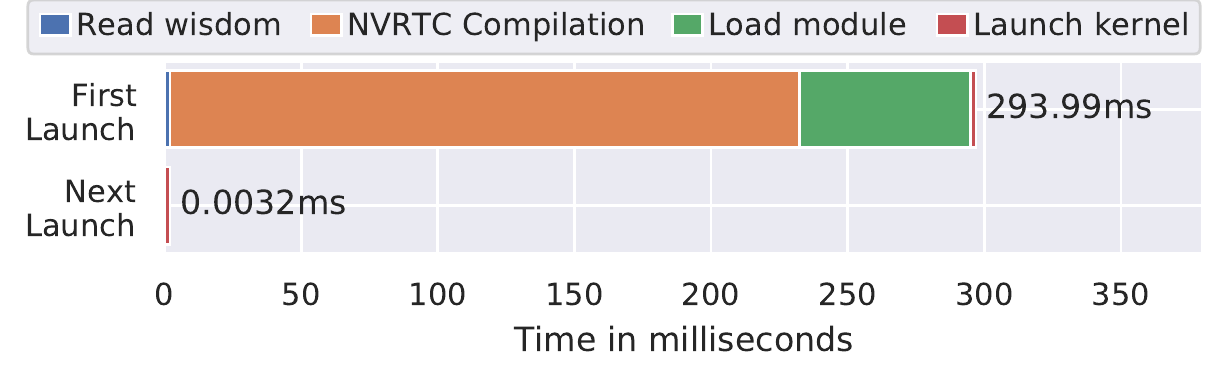}
\caption{Average time required for the first and subsequent launches of a kernel using Kernel Launcher. See Section~\ref{sec:evaluation_overhead} for explanation.}
\label{fig:evaluation_overhead}
\end{figure}

For each kernel, Kernel Launcher introduces some overhead on the first launch since the kernel's source code must compiled dynamically at run time.
We found that the first kernel call takes, on average, $294$ ms for our benchmarks.
Figure~\ref{fig:evaluation_overhead} shows a breakdown of this time.
There are four sources of overhead: reading the wisdom file, runtime compilation using NVRTC (\texttt{nvrtcCompileProgram}), loading the compiled code into the GPU (\texttt{cuModuleLoad}), and scheduling the kernel onto the GPU (\texttt{cuLaunchKernel}).
The figure shows that the runtime compilation (NVRTC) is the most expensive stage and constitutes around 80\% of this overhead.

For subsequent launches, runtime compilation is not necessary since the compiled kernel is cached.
We found that subsequent kernel calls using Kernel Launcher take just 3 $\mu{}s$ on average.
This is comparable to the overhead of CUDA when launching a GPU kernel without the use of Kernel Launcher.

%\subsection{Kernel Launcher Overhead}
% Stijn: Maybe something on overhead of KL? something like time to compile kernel/load compile kernel? something like overhead in launching a kernel (should negligible)

% Ben: that would be very interesting, the question is I guess what to compare against, do you still have a baseline that doesn't use Kernel Launcher? Or could you measure overhead when comparing against using the default (is the default precompiled?) as a means to save work on performing this comparison

\section{Conclusion}\label{sec:conclusion}

% re-re-re-explain what we did
In this work, we have presented Kernel Launcher: A C++ library that eases the development of auto-tuned CUDA applications. 
Kernel Launcher introduces tunable kernel definitions that merge the tuning code into the host application, reducing redundancy and fragmentation in the source code.
Kernel Launcher introduces {\em capturing} tunable kernels by storing all the information required to tune the kernel during execution, fully automating the process of tuning kernels using Kernel Tuner.
Kernel Launcher then uses wisdom files to select and compile the optimal kernel configuration at runtime.
Finally, we have demonstrated that Kernel Launcher can be used to create optimal-performance portable CUDA applications, with a clean API that resembles the CUDA runtime API and with little extra work for application developers.

We have evaluated Kernel Launcher using MicroHH, a computational fluid dynamics code, on two GPUs, the A100 and A4000. Even though these GPUs are based on the same architecture (Nvidia Ampere), using a configuration tuned on one GPU, for a specific problem size and precision, on the other GPU only results in 70\%-77\% and 69\%-82\% (A100 to A4000), and 67\%-83\% and 50\%-77\% (A4000 to A100), of the potential performance for two kernels in MicroHH.
Moreover, using the performance portability metric, we have shown that tuning only for one scenario and then using this configuration in all scenarios may lead to worse overall performance compared to not tuning the code at all.
Using Kernel Launcher, the application always selects the optimal configuration to achieve 100\% of the potential performance in each scenario, thus achieving optimal performance portability.

This work is not without limitations. 
Applications for which the kernel problem size is unknown beforehand cannot be tuned at compile time. 
Kernel Launcher does support fuzzy matching to select kernels with the closest matching problem size, so if the distribution of problem sizes can be sampled Kernel Launcher can still be used, but performance may be suboptimal. 
Additionally, Kernel Launcher uses the problem size as the primary descriptive feature on which  configurations are selected for one GPU. For irregular problems, such as graphs or sparse matrices, performance may depend strongly on the data itself rather than its size. Kernel Launcher currently does not support selecting on features other than the problem size and GPU, and as such, performance may be suboptimal for irregular algorithms. 

% what have we learned

% explain in what sense the world will be a better place now

\section*{Acknowledgments}

The CORTEX project has received funding from the Dutch Research Council (NWO) in the framework of the NWA-ORC Call (file number NWA.1160.18.316). The ESiWACE2 and ESiWACE3 projects have received funding from EU Horizon 2020 (No 823988) and EuroHPC JU (No 101093054).

\bibliographystyle{IEEEtran}
\bibliography{lib}

\end{document}